\documentclass{elsart}					

 \usepackage{graphicx}

\usepackage{amssymb}

\begin{document}

\begin{frontmatter}




\title{Moon Shadow by Cosmic Rays under the Influence of Geomagnetic
Field and Search for Antiprotons at Multi-TeV Energies}



\author[Hirosaki]{M. Amenomori},
\author[Saitama]{S. Ayabe},
\author[IHEP]{X. J. Bi},
\author[Yokokoku]{D. Chen},
\author[Heibei]{S. W. Cui},
\author[Tibet]{Danzengluobu},
\author[IHEP]{L. K. Ding},
\author[Tibet]{X. H. Ding},
\author[Shandong]{C. F. Feng},
\author[IHEP]{Zhaoyang Feng},
\author[SWJ]{Z. Y. Feng},
\author[Yunnan]{X. Y. Gao},
\author[Yunnan]{Q. X. Geng},
\author[Tibet]{H. W. Guo},
\author[IHEP]{H. H. He},
\author[Shandong]{M. He},
\author[Kanagawa]{K. Hibino \corauthref{cor}},
\corauth[cor]{Corresponding author. }
\ead{hibino@icrr.u-tokyo.ac.jp}
\author[Utsunomiya]{N. Hotta},
\author[Tibet]{Haibing Hu},
\author[IHEP]{H. B. Hu},
\author[ICRR]{J. Huang},
\author[SWJ]{Q. Huang},
\author[SWJ]{H. Y. Jia},
\author[Konan]{F. Kajino},
\author[Shibaura]{K. Kasahara},
\author[Yokokoku]{Y. Katayose},
\author[Shinshu]{C. Kato},
\author[ICRR]{K. Kawata},
\author[Tibet]{Labaciren},
\author[CSSAR]{G. M. Le},
\author[Shandong]{A. F. Li},
\author[Shandong]{J. Y. Li},
\author[Tsinghua]{Y.-Q. Lou},
\author[IHEP]{H. Lu},
\author[IHEP]{S. L. Lu},
\author[Tibet]{X. R. Meng},
\author[Saitama,Waseda]{K. Mizutani},
\author[Yunnan]{J. Mu},
\author[Shinshu]{K. Munakata},
\author[Utsunomiya2]{A. Nagai},
\author[Hirosaki]{H. Nanjo},
\author[NII]{M. Nishizawa},
\author[ICRR]{M. Ohnishi},
\author[Air]{I. Ohta},
\author[Saitama]{H. Onuma},
\author[Kanagawa]{T. Ouchi},
\author[ICRR]{S. Ozawa},
\author[IHEP]{J. R. Ren},
\author[TMCA]{T. Saito},
\author[ICRR]{T. Y. Saito},
\author[Konan]{M. Sakata},
\author[ICRR]{T. K. Sako},
\author[Kanagawa]{T. Sasaki},
\author[Yokokoku]{M. Shibata},
\author[ICRR]{A. Shiomi},
\author[Kanagawa]{T. Shirai},
\author[Shonan]{H. Sugimoto},
\author[ICRR]{M. Takita},
\author[IHEP]{Y. H. Tan},
\author[Kanagawa]{N. Tateyama},
\author[Waseda]{S. Torii},
\author[RIKEN]{H. Tsuchiya},
\author[ICRR]{S. Udo},
\author[Yunnan]{B. S. Wang},
\author[IHEP]{H. Wang},
\author[ICRR]{X. Wang},
\author[Shandong]{Y. G. Wang},
\author[IHEP]{H. R. Wu},
\author[Shandong]{L. Xue},
\author[Konan]{Y. Yamamoto},
\author[ICRR]{C. T. Yan},
\author[Yunnan]{X. C. Yang},
\author[Shinshu2]{S. Yasue},
\author[CSSAR]{Z. H. Ye},
\author[SWJ]{G. C. Yu},
\author[Tibet]{A. F. Yuan},
\author[Kanagawa]{T. Yuda},
\author[IHEP]{H. M. Zhang},
\author[IHEP]{J. L. Zhang},
\author[Shandong]{N. J. Zhang},
\author[Shandong]{X. Y. Zhang},
\author[IHEP]{Y. Zhang},
\author[IHEP]{Yi Zhang},
\author[Tibet]{Zhaxisangzhu},
\author[SWJ]{X. X. Zhou} \\

\bigskip

(Tibet AS$\gamma$ Collaboration)\\

\bigskip

\address[Hirosaki]{Department of Physics, Hirosaki University, Hirosaki 036-8561, Japan}
\address[Saitama]{Department of Physics, Saitama University, Saitama 338-8570, Japan }
\address[IHEP]{Key Laboratory of Particle Astrophysics, Institute of High Energy Physics, Chinese Academy of Science, Beijing 100049, China}
\address[Yokokoku]{Faculty of Engineering, Yokohama National University, Yokohama 240-8501, Japan }
\address[Heibei]{Department of Physics, Hebei Normal University, Shijiazhuang 050016, China}
\address[Tibet]{Department of Mathematics and Physics, Tibet University, Lhasa 850000, China }
\address[Shandong]{Department of Physics, Shandong University, Jinan 250100, China }
\address[SWJ]{Institute of Modern Physics, South West Jiaotong University, Chengdu 610031, China }
\address[Yunnan]{Department of Physics, Yunnan University, Kunming 650091, China }
\address[Kanagawa]{Faculty of Engineering, Kanagawa University, Yokohama 221-8686, Japan}
\address[Utsunomiya]{Faculty of Education, Utsunomiya University, Utsunomiya 321-8505, Japan}
\address[ICRR]{Institute for Cosmic Ray Research, University of Tokyo, Kashiwa 277-8582, Japan }
\address[Konan]{Department of Physics, Konan University, Kobe 658-8501, Japan}
\address[Shibaura]{Faculty of Systems Engineering, Shibaura Institute of Technology, Saitama 330-8570, Japan}
\address[Shinshu]{Department of Physics, Shinshu University, Matsumoto 390-8621, Japan}
\address[CSSAR]{Center of Space Science and Application Research, Chinese Academy of Sciences, Beijing 100080, C
hina}
\address[Tsinghua]{Physics Department and Tsinghua Center for Astrophysics,Tsinghua University, Beijing 100084, China}
\address[Waseda]{Advanced Research Institute for Science and Engineering, Waseda University, Tokyo 169-8555, Japan}
\address[NII]{National Institute for Informatics, Tokyo 101-8430, Japan}
\address[Air]{Tochigi Study Center, University of the Air, Utsunomiya 321-0943, Japan}
\address[Utsunomiya2]{Advanced Media Network Center, Utsunomiya University, Utsunomiya 321-8585, Japan}
\address[TMCA]{Tokyo Metropolitan College of Aeronautical Engineering, Tokyo 116-0003, Japan}
\address[Shonan]{Shonan Institute of Technology, Fujisawa 251-8511, Japan}
\address[RIKEN]{RIKEN, Wako 351-0198, Japan}
\address[Shinshu2]{School of General Education, Shinshu University, Matsumoto 390-8621, Japan}
\author{}

\address{}

\begin{abstract}
We have observed the shadowing of galactic cosmic ray flux in the direction of the moon,
the so-called moon shadow, using the Tibet-III air shower array operating 
at Yangbajing (4300 m a.s.l.) in Tibet since 1999.
 Almost all cosmic rays are positively charged; for that reason, they
are bent by the geomagnetic field, thereby shifting the moon shadow westward. 
The cosmic rays will also produce an additional shadow in the eastward direction of the moon if cosmic rays contain negatively charged particles, such as antiprotons, with some fraction. 
We selected $1.5 \times 10^{10}$ air shower events with energy beyond about 3 TeV from the
dataset observed by the Tibet-III air shower array and detected the moon shadow at $\sim 40 \sigma$ level.
The center of the moon was detected in the direction away from the apparent center of the moon
 by 0.23$^\circ$ to the west. Based on these data and a full Monte Carlo simulation, 
we searched for the existence of the shadow produced by antiprotons at the multi-TeV energy region.
No evidence of the existence of antiprotons was found in this energy region. 
We obtained the 90\% confidence level upper limit of the flux ratio of antiprotons to 
protons as 7\% at multi-TeV energies.
\end{abstract}

\end{frontmatter}

\section{Introduction}
\label{Introduction}

Observation of antiproton abundance in cosmic ray flux raises the
possibility of a baryon-symmetric universe and the propagation of cosmic rays in 
interstellar space. Recent measurements of antiproton flux, however, appear to be
almost within the conventional cosmic ray physics in which antiprotons are produced
as secondary particles of cosmic ray interactions with interstellar gas.
For example, antiprotons are produced mainly by collisions of 
cosmic-ray protons with interstellar hydrogen gas as 
$ p + p \longrightarrow \overline{p} + p + p + p $.
Accelerator experiments measured the production rate ($\bar{p}/p$) of antiprotons to protons to 
be on the order of $10^{-3} $ at energies greater than 10 GeV. 

Recently, measurements of absolute flux of antiprotons below a few GeV
were carried out using the magnet spectrometer equipped with a track 
detector to identify the charge and momentum of each incident particle 
 \cite{BASIN1999,HEAT2001,CAPRICE2001,BESS2002}.
 Among these, the CAPRICE2 experiment 
extended the energy range of the spectrum up to about 50 GeV.
It is obvious that accurate measurements of $\bar{p}$ flux are key to testing current
propagation models of cosmic rays in the Galaxy. Of course, a $\bar{p}$ ^^ ^^ excess" from the
reliable propagation model may lead to discover possible sources of primary antiprotons
such as dark matter annihilation and evaporation of primordial black holes.
Strong et al. \cite{STRONG2007} made a detailed calculation of the antiproton flux, diffuse gamma rays  and other
cosmic ray fluxes and compared it with the recent BESS results available in the energy region below a few GeV \cite{BESS2002}.
They found that the conventional local cosmic ray measurements, simple energy dependence of the
diffusion coefficient, and uniform cosmic ray source spectra through the
Galaxy fail to reproduce simultaneously both the secondary to
primary nuclei ratio and antiproton flux in this energy region.

The experiments mentioned above 
 also present the flux ratio, $\overline{p}/p$,
 of antiprotons to protons
in the cosmic rays. It seems that the observed results are within the
scope of the prediction of the standard leaky box model \cite{SIMON1998}. 
Some results \cite{BASIN1999,CAPRICE2001}, however, show
a tendency that the $\overline{p}/p$ ratio increases with increasing 
primary energy, although the amount of data is insufficient.

In the simple leaky box model \cite{MENEG1971}, the ratio of antiproton flux $f_{\bar{p}}(E_{\bar{p}})$ 
to proton flux $f_p(E_p)$ is calculated as \cite{GAIS1990}
\begin{equation}
   \overline{p} / p \equiv f_{\bar{p}}(E_{\bar{p}})/f_p(E_p) \propto \frac{\lambda_{esc}(E_p)}{\lambda_N}Z_{N\bar{N}},
\end{equation}
where $\lambda_{esc} \equiv \rho \beta c \tau_{esc}$ ($c$ is the light velocity) is 
the mean amount of matter (density $\rho$) traversed by a nucleon 
of velocity $\beta c$, $\tau_{esc}$ is the mean time spent by the cosmic rays in the confinement space, and
 $\lambda_N$ is the mean free path of the nucleon in the interstellar matter. In addition, $Z_{N\bar{N}}$ is the production rate of antinucleons by nucleon - nucleon (hydrogen atom) interactions, which
is almost constant when Feynman scaling holds, while weakly depending
 on the spectral index of cosmic rays. 
According to the HEAO-3 data on the B/C ratio in the primary cosmic rays up to about 50 GeV/n, 
$\lambda_{esc} \propto E^{-\delta}$ and $\delta \sim 0.6$ \cite{ENGEL1990}. Because the mean free path of the nucleon,
$\lambda_N$, in this energy region is almost constant, the flux ratio of
$\overline{p}/p$ should decrease with
increasing primary energy. At high energies greater than 100 GeV, however, a value of $\delta \sim 0.3$
 is expected because the cosmic ray anisotropy is less than $10^{-3}$
in the energy region below 100 TeV. Actually, $\delta$ should 
take a value of 1/3 when the magnetic field in interstellar space has 
a Kolmogorov-type spectrum.

A closed galaxy model \cite{PETERS1977} predicts a considerable enhancement of secondary antiprotons.
In this model, the cosmic ray flux observed locally in the solar system comprises young and
old components. Cosmic rays in the young component show almost identical behavior to
that discussed above, whereas the
old component is contained in the galaxy halo. The old component of cosmic rays, however, 
consists only of protons because heavier nuclei in the old component
are completely broken up during their long confinement time in the halo. 
Consequently, the value of $\lambda_{esc}$ for protons in the old component 
becomes very large.
This model can engender a considerable enhancement of secondary antiprotons because of the existence of
such old protons \cite{PROT1981}.

Until now, almost all measurements of antiprotons are limited in the energy region below a few GeV
 because of its extremely low flux and weight limit of the instrument on board the balloon.
In this energy region, the interpretation of observed spectra is not free from uncertainties 
about the secondary antiproton production because of the nuclear threshold effect and because of uncertainties in the solar modulation
effect. For those reasons, it is still difficult to distinguish which model is the best for explaining the
 antiproton spectrum. Considering the latest research, the possibility of the existence of
antiprotons in the TeV energy region is slight, but we cannot exclude an exotic model
such as an anti-galaxy production model yet (see the review paper by Stephen \& Golden \cite{STEPHEN1987}).
Therefore, it will be important to measure $\overline{p}/p$ ratio in 
the TeV energy region.
The results of such measurements might enable us to investigate the possible production of antiprotons from
dark matter at high energies \cite{BARRAU2005}.

Urban et al. \cite{Urban} first discussed the possibility of observing antiprotons using the
moon shadow; the first observation was done by the Tibet-I experiment \cite{Amenomori1992} 
by observing the sun shadow, giving an upper limit of the $\overline{p} / p$ ratio at 10 TeV \cite{Tibet1995}.
The Tibet air shower experiment has been observing the shadowing of galactic cosmic rays continuously by
the moon and sun since 1990 \cite{Amenomori1993}. 
We began construction of the Tibet-III air shower array in 1999 by increasing the
number of detectors; we completed it in Fall of 2003. The present Tibet-III air shower array 
consists of 789 scintillation counters of 0.5 m$^2$, each of which is placed in a 7.5 m square grid, 
covering a total area of about 37000 m$^2$ \cite{Tibet-Performance}.

In this paper, we present the result on the search for antiprotons based on the moon shadow 
observed during the period from 1999 through 2004 using the Tibet-III air shower array.

\section{Experiment} 
\label{Experiment}

The Tibet-III air shower array used in this experiment was constructed in 1999
 at Yangbajing (4300 m a.s.l.) in Tibet and operated until 2004.
The array, corresponding to the inner part of the full-scale Tibet-III air shower array,
 consists of 533 scintillation counters covering 22050 m$^2$ \cite{Amenomori2003,Science2006}.
The mode energy of detected events is about 3 TeV for proton-induced showers 
and the angular resolution is 0.9$^\circ$. The systematic error of the energy determination
of primary particles and systematic pointing error of the array have been well
 calibrated by comparing the observed displacement of the moon shadow 
because of the geomagnetic field with the Monte Carlo simulation, as discussed 
in the paper \cite{Kawata}.

With the advent of this array, we observed $ 6.1 \times 10^{10} $ events during 
the period from November 1999 through December 
2004 (1041 live days in total). These events were selected by imposing the following conditions:
1) Each shower must fire four or more counters recording 1.25 or more particles; 
2) all of fired counters or eight of nine fired counter which recorded the
highest particle density must be inside the fiducial area; and 
3) the zenith angle of the arrival direction must be less than $ 40^{\circ}$.
After the these selections, $1.5 \times 10^{10} $ events remained for further analyses.

\section{Simulation}
\label{Simulation}

We have done a detailed Monte Carlo simulation to estimate the flux of antiprotons from 
the shape of the moon shadow by cosmic rays under the influence of geomagnetic field. A deviation of
the observed shadow from the simulated one based on the normal primary cosmic rays
 (without antiparticles) would give the proof  of  the existence of antiprotons pouring on the Earth. 
For the normal primary cosmic rays, we used a best-fit curve for each component based on
the experimental data measured by direct observations on board the balloon and satellite (for 
experimental data, see the paper by Gaisser et al. \cite{GAISSER2001} and references therein).

We used  the CORSIKA Ver. 6.200 code \cite{CORSIKA} with the QGSJET interaction model 
 for generation of air showers in the atmosphere. All secondary particles produced by
a primary particle were traced until their energies becomes 1 MeV in the atmosphere.
Simulated air-shower events were then input to the detector with the same detector configuration
as the Tibet-III air shower array using the Epics uv8.00 code \cite{EPICS}
 to calculate the energy deposit of these shower particles and were stored in the disk with
the same format as the experimental data. 

For the geomagnetic field, we used the Virtual Dipole Moment (VDM) model \cite{VDM} in which
the dipole moment at altitude $> 600$ km is taken to be 
 $ 8.07 \times 10^{25} ~\rm{G \cdot cm^{3}} $. 
We also made the simulation based on the International Geomagnetic Reference Field (IGRF) model \cite{IGRF} 
 to compare with that based on the VDM model.
The difference of the deflection of particle trajectries between the two models is then estimated to be within  
 5 $\sim$ 7\% in the part of the sky where the moon is visible by  the Tibet-III air shower array.
We found, however, that the  IGRF model gives almost the same result as the VDM model
for the actual moon shadow simulation in which the angular resolution of the array is
taken into account.  Hence, we adopted the results obtained based on the VDM model in this paper.

The simulation for the moon shadow was performed through the following steps:

\begin{enumerate}
\item Primary particles in the energy range from 0.3 TeV to 1000 TeV are thrown isotropically toward the observation site
on the top of the atmosphere along the revolution orbit of the moon. Air shower events generated by these particles in the atmosphere are then collected within the circle of 300 m radius from the center of an air shower array.
This array has the same structure and detector response as the Tibet-III air shower array.

\item  All air shower events triggered by the air shower array 
 are analyzed in the same way as the experimental data to estimate the size
 and the arrival direction of each event.

\item An opposite charge is then assigned to the primary particle of each analyzed event;
and these particles (anti-particles) are again shot back uniformly in 
the $\pm 10^{\circ} \times \pm 10^{\circ}$ direction centering
on the moon: they originated from the
air shower array at Yangbajing and tracked through the geomagnetic field up to the moon.

\item We select only the events hitting the moon; these selected events are
again traced back from the moon to the Earth after reversing
the charge of each primary particle. The final direction of each cosmic ray is determined 
by smearing the difference between the moon direction and the estimated air shower 
direction where the angular resolution of the array is taken
into account. 
\end{enumerate}

Using this method, we can simulate the moon shadow that is 
capable of direct comparison with the observed data. 

\section{Data Analysis}
\label{Analysis}

For analysis of the moon shadow, an on-source event is defined as an event which
falls on the $ \pm 5^{\circ} \times \pm 5^{\circ} $ square window centered 
at the moon direction.
The background is estimated using the events coming from the
following eight off-source windows of the same size as on-source.
These positions are located on the same zenith angle as the moon, but
separated by $\pm 5^\circ, \pm 10^\circ, \pm 15^\circ$, and $\pm 20^\circ$ in the
azimuth angle. 
The number of events averaged over these eight off-source windows
was taken to be the background \cite{Amenomori1993}. 

\begin{figure}[hbtp]
  \begin{center}
    \includegraphics*[width=9cm]{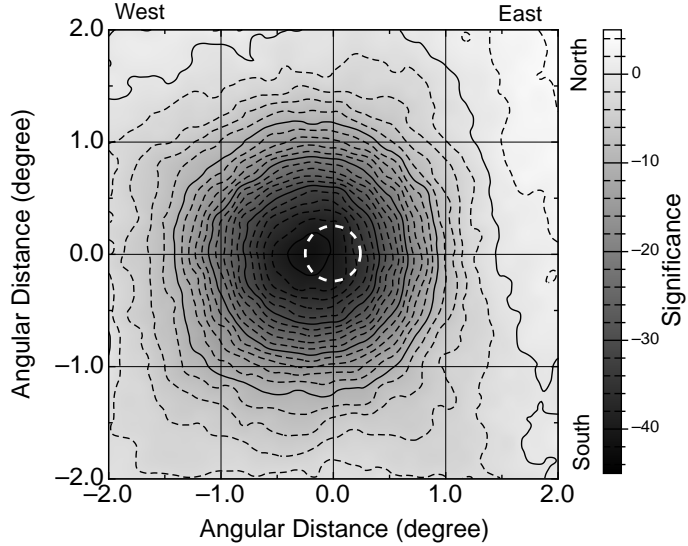}
  \caption{Significance map of the deficit event densities 
  observed with the Tibet-III air shower array for 1041 live days. 
  The origin of the coordinate is taken in the direction of the moon.
  The white dashed circle denotes the apparent size and position of the moon.
  A right-hand side scale expresses the level of significance of the deficit event 
  density in terms of the standard deviation $\sigma$.
  The significance at the central peak of the shadow is about
40$\sigma$ on the two-dimensional analysis \cite{Amenomori1993}. }
\label{fig:significance}
\end{center}
\end{figure}

Shown in Fig. \ref{fig:significance} is the moon shadow observed using the
events obtained for 1041 live days with the Tibet-III air shower array.
The significance at the peak of the shadow is estimated as about 40$\sigma$ level
and the shadow is shifted westward by about $0.23^\circ$.
The mode energy of observed events is estimated as 3 TeV for 
proton-induced showers. Note that a 1 TeV proton is shifted westward by
$1.6^\circ$ at Yangbajing. This means that  a major effect of
 antiprotons should appear in the eastward angular distance smaller than about 0.5$^\circ$
 on the moon shadow observed with the Tibet-III air shower array. 
 The systematic error of the energy estimation of
primary particle is estimated to be $\pm$ 8\% level and the systematic
pointing error of the array is smaller than $0.011^\circ$ \cite{Kawata}.
Using these basic data and simulation results, we discuss the flux of antiprotons
in the next section.

\section{Results and Discussion}
\label{Results}

\begin{figure}
  \begin{center}
    \includegraphics*[width=9cm]{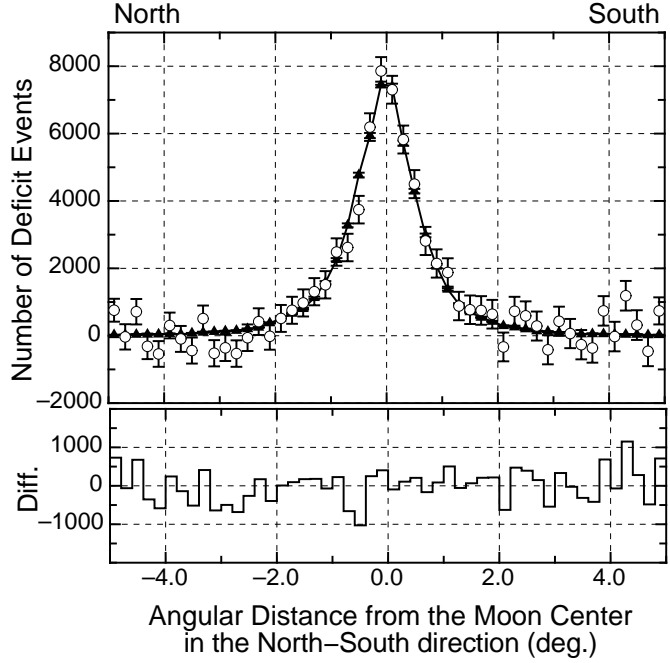}
  \caption{One-dimensional distribution of the observed deficit events around 
the center of the moon in the north-south direction. 
The vertical axis represents the number
of deficit events inside the range of $\pm 1^{\circ}$ in the east-west direction.
The open circles denote the experimental data and the
filled triangles denote the MC data. The solid line represents
a best fit curve according to the function $f_1(\theta)$.
Each error bar denotes $\pm 1 \sigma$.
The histogram of the following figure shows the residual of the observed deficit event
distribution to the function $f_1(\theta)$.}
\label{fig:comp_exp_sim_NS}
\end{center}
\end{figure}

Figure \ref{fig:comp_exp_sim_NS} shows the one-dimensional distribution
of the observed deficit events around the moon in the north-south direction, where the vertical
axis represents the number of deficit events inside the range of $\pm 1^\circ$ in the east-west direction.
The origin of the coordinate is taken to indicate the apparent direction 
of the moon. The position of the shadow center is almost consistent with
that of the moon because the east-west component of the geomagnetic
field is negligibly small at Yangbajing latitude, so that the cosmic rays are not deflected in the
north-south direction. As this figure shows, our data show good agreement 
with the MC simulation ($\chi^2/d.o.f. \sim 1.15$).
 The peak position of the distribution in the north-south direction is estimated as 
$0.0048^\circ \pm 0.011^\circ$.

\begin{figure}
  \begin{center}
    \includegraphics*[width=9cm]{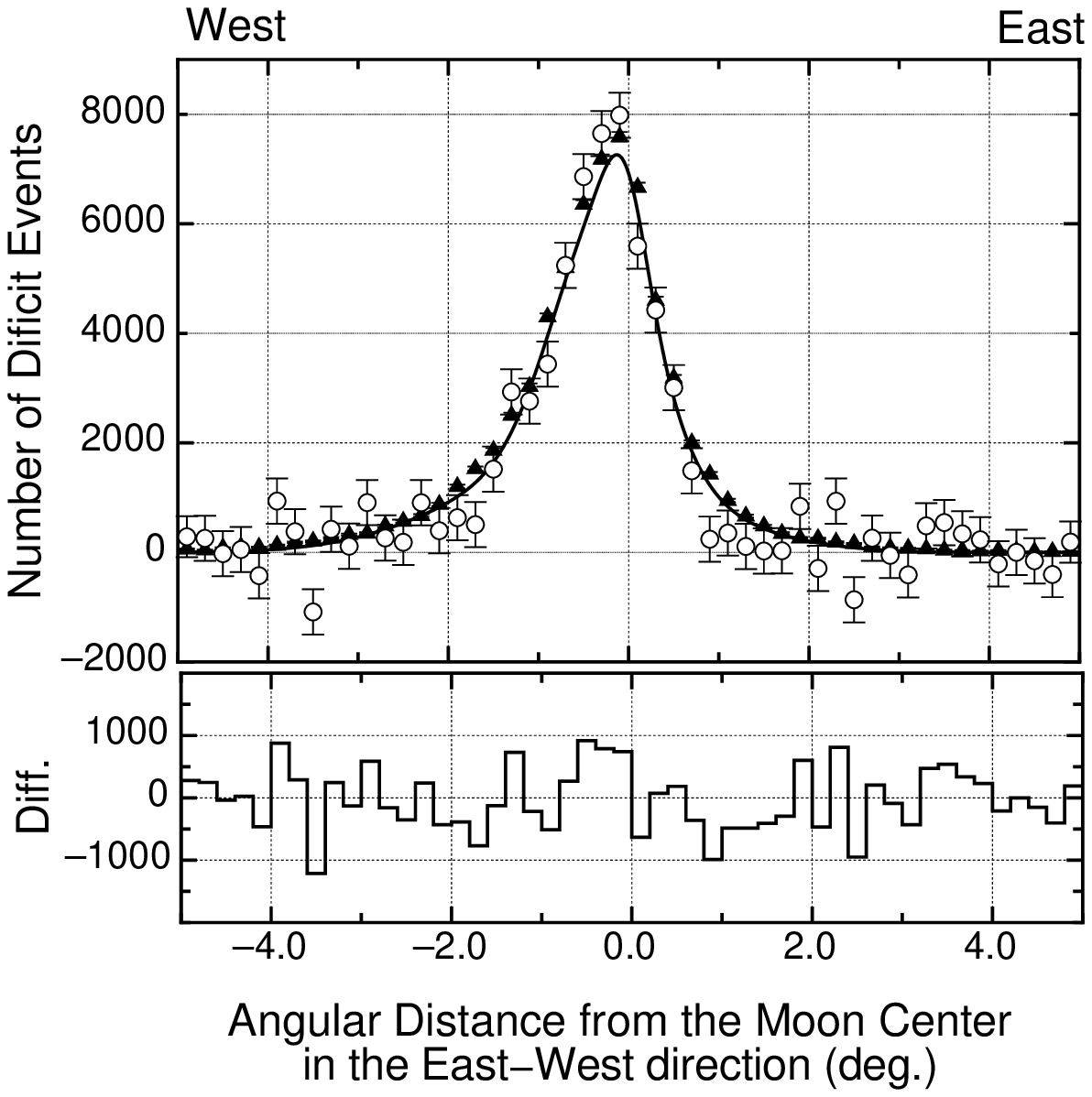}
  \caption{One-dimensional distribution of the observed deficit events around the
center of the moon in the east-west direction.
 The vertical axis represents the number of deficit events inside the range of $\pm 1^{\circ}$ 
in the north-south direction. 
 The open circles denote the experimental data and the filled triangles
denote the MC data without antiprotons. The solid line is a best-fit curve according to the
function $f_3(\theta)$.
Each error bar denotes $\pm 1 \sigma$.
The histogram of the following figure shows the residual of the observed deficit event
distribution to the function $f_3(\theta)$.}
\label{fig:comp_exp_sim}
\end{center}
\end{figure}

Figure \ref{fig:comp_exp_sim} shows the one-dimensional distribution 
 of the observed deficit events around the moon in the east-west direction, where the vertical
axis represents the number of deficit events inside the range of $\pm 1^\circ$ in the north-south direction.
The observed moon shadow is shifted westward 
by $ 0.23^{\circ} $, which is consistent with the MC simulation. 
There exists, however, no evidence 
indicating deficits of cosmic rays at the opposite position 
around $\theta = 0.23^\circ$ in the eastward direction,
corresponding to the particles with negative charge (anti-matter)
 such as $\overline{p}, \overline{He}, \overline{C}$,.., and $\overline{Fe}$, if any.

Next, to evaluate the contribution of antiprotons to the moon shadow,
we obtained two kinds of moon shadow that were cast by all cosmic rays and protons, respectively.
We assume that the deficit event distribution in each shadow can be expressed as
a superposition of Gaussian functions.
It is found that four Gaussian functions are adequate for fitting to both distributions 
in the angular distance smaller than about 5$^\circ$ from the center of the moon.

The moon shadow by all cosmic rays is then expressed as
\begin{eqnarray}
f_{1}(\theta) = \sum_{i=1}^{4} A_{all,i} \exp\left(- \frac{4 \ln 2 \times (\theta-M_{all,i})^{2}}{\sigma_{all,i}^{2}}\right).
\end{eqnarray}

The moon shadow by protons is also expressed as
\begin{eqnarray}
f_{2}(\theta) = \sum_{i=1}^{4} A_{p,i} \exp\left(- \frac{4 \ln 2 \times (\theta-M_{p,i})^{2}}{\sigma_{p,i}^{2}}\right),
\end{eqnarray}
where $ \theta$ is the angular distance from the moon center
 in the west-east direction and $A, M$ and $\sigma$ are the fitting parameters for the
 distribution function, respectively, for all cosmic rays and protons.

On the other hand, the observed moon shadow should be 
expressed by the following function $ f_{3}(\theta) $ 

\begin{equation}
f_{3}(\theta) = a f_{1}(\theta) + b f_{2}(-\theta),
\end{equation}
where the first term represents the deficit in cosmic rays and the
second term represents the deficit in antiprotons. 
 From our simulation, the fraction of protons to all cosmic rays is estimated as $ 62 \pm 1\% $ of all 
cosmic rays, so the ratio $ b/0.62a $ corresponds to 
the $\overline{p}/p$ ratio. 
As shown in Fig.~\ref{fig:comp_exp_sim}, the observed deficit events
can be fitted by the function $ f_{3}(\theta) $ with the
parameters of $a = 1.64 \pm 0.10$ and $b = -0.10 \pm 0.08$ 
($\chi^2/d.o.f. \sim 1.50$),  where the fitting was made under the boundary condition 
that the integral of $af_1(\theta)$ should be smaller than the total number of observed events.

\begin{figure}
 \begin{center}
    \includegraphics[width=9 cm]{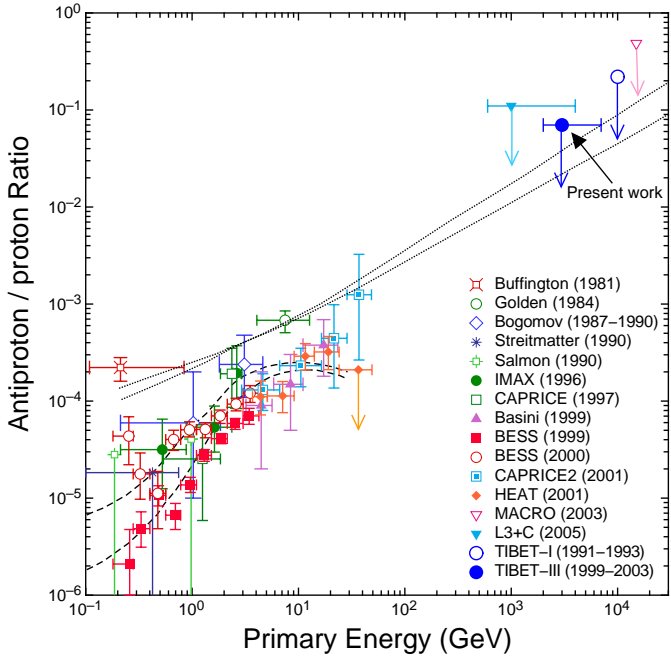}
 \caption{The antiproton/proton ratio at the top of the atmosphere. 
The experimental results are as follows: 
Buffington (1981) \cite{Buffington},
Golden (1984) \cite{Golden},
Bogomolov (1987-1990) \cite{Bogomolov},
Salomon (1990) \cite{Salomon},
IMAX \cite{IMAX},
CAPRICE (1997) \cite{CAPRICE1997}, 
Basini (1999) \cite{BASIN1999},
BESS (1999) and BESS (2000) \cite{BESS2002}, 
CAPRICE2 (2001) \cite{CAPRICE2001},
HEAT (2001) \cite{HEAT2001},
L3+C (2005) \cite{L3C2005},
Streitmatter (1990) \cite{Streitmatter},
MACRO (2003) \cite{MACRO},
L3+C (2005) \cite{L3C2005},
and opened circles are those by TIBET-I (1990--1993) \cite{Tibet1995}. 
The present result by TIBET-III (1999--2004) is denoted by the filled circle.
The dashed lines, upper and lower limit, are the calculation based on
the leaky box model by Simon, Molnar \& Roesler \cite{SIMON1998}.
The dotted lines are for a model that includes both leaky box  and
  the presence of antistars. The rigidity dependent confinement of
cosmic rays in the Galaxy is assumed to be $\propto R^{-\delta}$, and two curves $A$ and $B$ are
 the cases of $\delta = 0.7$ and 0.6, respectively \cite{STEPHEN1987}.
}
\label{fig:PbarPratio}
\end{center}
\end{figure}

To calculate an upper limit of the fraction of antiprotons, 
the confidence interval of the parameter $b$ is estimated using 
Feldman and Cousins Statistics \cite{Feldman} because parameter $b$ is negative.
The upper limit of the $\overline{p}/p$ ratio at the 90\% confidence level was then
obtained as $0.07 $. The present result is plotted in Fig. \ref{fig:PbarPratio},
 together with other results. 

In this figure, four experimental data are plotted at multi-TeV energies.
Among them, two are from the Tibet air-shower experiment using the 
moon shadow (present work) and the sun shadow (Tibet-I \cite{Tibet1995}).
The present result gives the most stringent constraint on the $\overline{p}/p$
 ratio compared to other data.

In conclusion, we showed that the moon shadow observation can provide
unique data on the $\overline{p}/p$ ratio at the multi-TeV region.
A further observation of the moon shadow and a fine tuning of the MC simulation
will give a more stringent constraint to the flux ratio of antiprotons to
protons in the very near future. If these data are combined with the
new data that are obtainable in the energy region around 100 GeV
by the PAMELA satellite mission (launched on board the Russian 
satellite in 2006) \cite{PICOZZA2007} and by the forthcoming space
experiment AMS \cite{AMS2}, we might be able to get new information about
 the production of antiprotons in the TeV energy region.
 
\section{Acknowledgments}

The collaborative experiment of the Tibet Air Shower
Arrays has been performed under the auspices of the
Ministry of Science and Technology of China and the
Ministry of Foreign Affairs of Japan. This work was 
supported in part by Grants-in-Aid for Scientific Research
on Priority Areas (712) (MEXT) and by Scientific Research (JSPS) in Japan,
and by the Committee of the Natural Science
Foundation and by the Chinese Academy of
Sciences in China.



\end{document}